\documentstyle[sprocl,psfig]{article}

\def\Journal#1#2#3#4{{#1} {\bf #2}, #3 (#4)}

\def\NPB{{\em Nucl. Phys.} B}
\def\PLB{{\em Phys. Lett.}  B}

\def\PRD{{\em Phys. Rev.} D}
\def\ZPC{{\em Z. Phys.} C}

\def\be{\begin{equation}}
\def\ee{\end{equation}}
\def\bea{\begin{eqnarray}}
\def\eea{\end{eqnarray}}

\begin{document}

\title{HIGH-ENERGY SCATTERING AND DIFFRACTION:\\ THEORY
  SUMMARY\footnote{Talk given at the {\em 8th International Workshop
      on Deep-Inelastic Scattering (DIS2000)}, 25th-30th April 2000,
    Liverpool, England, to appear in the proceedings.}}

\author{A. HEBECKER\raisebox{1ex}{\footnotesize 1} and 
T. TEUBNER\raisebox{1ex}{\footnotesize 2}}

\address{\vspace*{0.3cm}\raisebox{1ex}{\footnotesize 1}
Institut f\"ur Theoretische Physik, Universit\"at Heidelberg, 
Philosophenweg 16,\\ D-60120 Heidelberg, Germany
\\E-mail: hebecker@thphys.uni-heidelberg.de} 

\address{\raisebox{1ex}{\footnotesize 2}
Institut f\"ur Theoretische Physik E, RWTH Aachen, D-52056 Aachen, 
Germany \\E-mail: teubner@physik.rwth-aachen.de}

\maketitle\abstracts{ New developments in the theory and phenomenology of 
high-energy scattering and diffraction that were presented and
discussed at DIS2000 are reviewed.}

\section{Introduction}
On the one hand, small-coupling perturbation theory has been successfully 
applied to a variety of QCD processes. Its validity is well-understood in 
situations where intermediate states with high virtualities dominate. On the 
other hand, lattice Monte Carlo simulations provide a powerful 
first-principles approach to study the low-energy characteristics of the 
theory, such as the spectrum of hadronic excitations. However, there is 
still no established method, derived from the Lagrangian of QCD, that 
describes the high-energy scattering of hadrons. The reason for this is the 
difficulty to combine non-perturbative effects with the fundamentally
Minkowskian physics in the high-energy limit. Thus, it can be argued that the 
high-energy limit represents one of the most interesting and difficult 
open problems in the theory of strong interactions. One obvious 
challenge is the derivation of the high-energy behaviour of hadronic cross 
sections, which are well-parametrised as $\ln^2s$ or $s^{0.08}$ (where 
$\sqrt{s}$ is the cms-energy of the collison), from the known microscopic 
theory. 

\begin{figure}[ht]
\begin{center}
\parbox[b]{10cm}{\psfig{figure=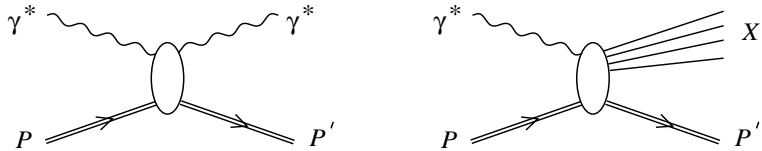,width=10cm}}\\
\end{center}
\vspace{-0.2cm}
\caption{Forward Compton scattering and diffractive electroproduction. 
\label{fig:dep}}
\end{figure}

Diffraction, and in particular the processes of hard diffraction discovered 
at the CERN S$p\bar{p}$S collider and studied in detail at HERA and the 
Tevatron, represent a powerful tool for the study of the high-energy limit 
of QCD. This is illustrated in Fig.~\ref{fig:dep}, where forward Compton 
scattering, equivalent to the process of deep-inelastic scattering (DIS), is 
compared to diffractive electroproduction. Obviously, the study of different 
diffractive final states $X$ provides a wealth of hadronic high-energy 
scattering data, taking us far beyond the well-known inclusive process
of DIS.

\section{New Approaches to the High-Energy Limit of QCD}
A fundamentally new approach to the high-energy limit of QCD has been 
advertised by Peschanski~\cite{pesch,jp}. The authors suggest using the 
AdS/CFT correspondence (also known as the Maldacena conjecture)~\cite{mal} 
to investigate high-energy scattering in non-Abelian gauge theories. 
AdS/CFT correspondence claims the equivalence of weakly coupled string 
theory in an Anti-de-Sitter (AdS) geometry with strongly coupled 
${\cal N}=4$ super Yang-Mills theory, which is a conformal field theory
(CFT), in 4-dimensional Minkowski space. Further, to make the connection 
with the realistic case of confining gauge theories, the authors use 
Witten's proposal~\cite{wit} that a confining gauge theory is dual to 
string theory in an AdS black hole background. In the gauge theory, the 
high-energy scattering of two dipoles can be calculated from the correlation 
function of two Wilson loops. Using AdS/CFT correspondence, the 
calculation of the latter can be reduced to a minimal surface problem 
in an AdS black hole background. The results obtained so far show 
reggeization with unit intercept~\cite{jp}. 

A very different unconventional approach to high energy scattering has 
been suggested by Kharzeev and Levin~\cite{khar,lev,kl}. They start from 
the leading order BFKL ladder diagram and emphasize the NLO contribution 
where the one-gluon rungs are replaced by pairs of gluons. Then, focussing 
on the soft region, these gluon pairs are replaced by pion pairs. The 
coupling to the vertical gluon lines of the ladder is fixed by employing 
the QCD anomaly relation 
\be
\theta_\mu^\mu=\frac{\beta(g)}{2g}\,F^{a\,\alpha\beta}F^a_{\alpha\beta}
\ee
and calculating the trace of the energy momentum tensor $\theta_\mu^\mu$ 
in terms of the pion degrees of freedom in chiral perturbation theory. 
After all ladder diagrams with two-pion rungs are summed, a 
soft-pomeron-like behaviour $\sim s^\Delta$ emerges. The intercept 
$\Delta=(1/48)\ln M_0^2/m_\pi^2$ comes out approximately right if the 
matching scale $M_0^2$, introduced by using chiral perturbation theory, is 
taken in the range $4\div 6$ GeV$^2$, in agreement with sum rule analyses. 

As emphasized by Kharzeev, diffractive glueball production provides an 
interesting testing ground for this new picture of the pomeron~\cite{khar}.

\section{$\gamma^*$-$\gamma^*$ Scattering at High Energy}
The total cross section of two highly virtual photons represents a 
unique testing ground for perturbative methods in high-energy scattering 
because the underlying process is the interaction of two small colour 
dipoles. The process is expected to include a kinematical region where the 
BFKL summation techiques are applicable. 

Recent progress relevant to NLO BFKL calculations~\cite{nlo} was discussed 
by Lipatov, who emphasized the enormous simplifications of the NLO BFKL 
kernel arising in ${\cal N}=4$ SUSY QCD and possible close relations between 
BFKL and DGLAP~\cite{lip,koli}. Furthermore, Lipatov noted the good 
description of $\gamma^*\gamma^*$ data in NLO BFKL achieved by using a 
non-Abelian physical renormalization scheme together with BLM scale 
fixing~\cite{blm}. However, other methods to modify the naive NLO 
corrections to BFKL, which are extremely large, do also exist~\cite{mnlo}. 

The problems of the BFKL method justify the attempt to account for the 
data, which lies far above the Born term prediction, by other means. 
Naftali~\cite{naf,glmn} presented a calculation taking into account hard-soft 
and soft-soft contributions, which are also present in $\gamma^*$-$\gamma^*$ 
processes. Although significant enhancements were found, they are not 
sufficient to account for the data when both photon virtualities are 
large. Donnachie reviewed the recent phenomenological approach of the 
`two pomerons'~\cite{don,dl}, which includes the well-known reggeon and
soft pomeron trajectories and a phenomenological hard pomeron with an 
intercept $\sim 0.44$. This approach describes successfully 
$\gamma$-$\gamma$ and $\gamma^*$-$\gamma$ cross sections, but is below 
the data in the $\gamma^*$-$\gamma^*$ case.

\section{Diffractive Electroproduction}
A large part of the diffractive data at HERA can be characterized by the 
diffractive structure function $F_2^D$, which describes the process 
$\gamma^*p\to p'X$ (cf.~Fig.~\ref{fig:dep}). In addition to the conventional 
kinematic variables of DIS, $Q^2$ and $x=x_{\mbox{\scriptsize Bj}}$, the 
process is characterized by $M$, the mass of the diffractive final state 
$X$. Alternatively, the variables $\beta=Q^2/(Q^2+M^2)$ or $\xi=x_{I\!\!P}=
x/\beta$ can be used. Now, $F_2^{D(3)}(x,Q^2,\xi)$ is defined precisely as 
$F_2(x,Q^2)$, but on the basis of a cross section that is differential
in $\xi$ as well as in $x$ and $Q^2$. Elastic vector meson production, to 
be discussed in more detail below, is obtained by appropriately specifying 
the final state $X$. For recent theoretical reviews of diffractive DIS see 
refs.~\cite{rev1,rev2}. 

Diffraction occurs if the hadronic fluctuation of the incoming virtual 
photon scatters off the proton without destroying its colour neutrality. 
At leading order, the fluctuation is a $q\bar{q}$ pair, and its 
interaction can be parametrized by the dipole cross section $\sigma(\rho)$, 
where $\rho$ is the transverse size of the dipole~\cite{nz}. A QCD-improved 
parametrization of the dipole cross section which carefully implements its 
relation to the gluon distribution in the region of small $\rho$ and avoids 
unitarity violations associated with the strong growth of the gluon 
distribution at small $x$ was presented by McDermott~\cite{mcd,dfgs}. 

Diffractive parton distributions~\cite{dpd}, denoted here by $df^D_i/d\xi$, 
characterize the probability of finding a parton in the proton under the 
condition that the proton remains intact. In this framework, which is 
firmly rooted in perturbative QCD, the diffractive cross section reads 
\be
\frac{d\sigma(x,Q^2,\xi)^{\gamma^*p\to p'X}}{d\xi}=\sum_i\int_x^\xi dy\,
\hat{\sigma}(x,Q^2,y)^{\gamma^*i}\left(\frac{df^D_i(y,\xi)}{d\xi}\right)\, ,
\ee
where $\hat{\sigma}(x,Q^2,y)^{\gamma^*i}$ is the total cross section for 
the scattering of a virtual photon characterized by $x$ and $Q^2$ and a 
parton of type $i$ carrying a fraction $y$ of the proton momentum. 

Royon presented a parametrization of $F_2^D$ as well as a QCD fit based 
on the DGLAP evolution of diffractive parton distributions~\cite{roy,br}. 
A novel feature of this fit is the subtraction of higher twist contaminations 
at large $\beta$. It is interesting that the famous `peaked' gluon of 
previous H1 analyses~\cite{h1} seems to be disfavoured.

Sch\"afer~\cite{schaefer} discussed results for $F_2^{D(3)}$ at small
$\beta$, obtained in the colour-dipole Regge-expansion approach and
stressed the relevance of unitarity corrections for the $\xi$ dependence.

Goulianos has suggested a simple parametrization~\cite{gou} of the
$F_2^D$ data at HERA, which is based on the ansatz
$d^3\sigma/d\xi\,dx\,dQ^2\sim F_2^h(x,Q^2)/x/\xi^{1+\epsilon}$ with a
`hard' structure function $F_2^h$. 

An important new result concerning the charm contribution to $F_2^D$ was 
presented by Bartels~\cite{bar}. At leading order, diffractive charm 
production is realized by $c\bar{c}$ and $c\bar{c}g$ final states. Except 
for the large-$\beta$ region, the latter component dominates because it 
allows for soft colour-singlet exchange. The new results presented by 
Bartels extend previous calculations of $c\bar{c}g$ production, where the 
$p_\perp$ of the gluon was assumed to be much smaller than the $p_\perp$ 
of the quarks (strong $p_\perp$ ordering), to general kinematic
configurations excluding, however, the case of soft gluons.

\section{Diffraction at Hadron-Hadron Colliders}
A frequently discussed issue in hard diffractive processes where either 
one or both colliding hadrons remain intact is the question whether a 
simple connection with the partonic description of diffractive DIS can
be found. As emphasized by Royon, the hadronic data undershoots HERA based 
expectations by a large factor~\cite{roy,whit}, which is indeed expected 
from the simple geometrical picture of the collision of two extended soft 
objects. Thus, a fundamentally new theoretical approach to this type of 
hadronic processes appears to be necessary. 
Timneanu~\cite{tim}
reported the successful description of both HERA and Tevatron gap events by
using a Monte Carlo implementation of a Soft-Colour-Interaction model
based on the generalized area law.  Also, as presented in the talk by
Cox~\cite{cox,cfl}, HERA and Tevatron data characterized by a gap
between two jets can be described by LLA BFKL within the HERWIG Monte
Carlo, if a fixed $\alpha_S = 0.17$ is adopted and if multiple
scattering for the underlying event is taken into account. 
The interesting process of Higgs (or dijet) production in double rapidity
gap events was discussed by Khoze~\cite{kmr}.  He presented refined
calculations in a perturbative approach, which, however, lead to cross
sections considerably smaller than those predicted by some
non-perturbative models.  
Close explained~\cite{close} how the pomeron
can be studied in hadron collisions at low momentum transfer by
measuring the $\phi$ and $t$ dependence for different ($J^{PC} =
0^{\pm +}, 1^{++}, 2^{++}$) mesons. 

\section{Elastic Meson Production}
Elastic vector meson production $\gamma^* p \to V p$ is a rich field,
both theoretically and exprimentally.  For large photon virtualities
$Q^2$ and/or heavy mesons (with large mass $M_V$) this process
constitutes a nice laboratory to study diffractive hard scattering.
At small $\xi$ the production amplitude factorizes in the fluctuation
of the virtual photon, the elastic scattering of the $q\bar q$ (or $q\bar q
g, \ldots$ in higher orders) off the proton, and the formation of the
final state (vector) meson ($V$):  ${\cal A}(\gamma^* p \to V p) =
\psi_{q\bar q}^{\gamma} \otimes A_{q\bar q + p} \otimes \psi_{q\bar
  q}^V$.  At leading order the diffractive (colourless) exchange is 
realized by a pair of gluons~\cite{rys}. In the usual
collinear factorization approach the amplitude $A_{q\bar q + p}$ is
then proportional to the gluon density in the proton $\xi g(\xi,
\mu^2)$ at some effective scale $\mu^2 \sim (Q^2 + M_V^2)/4$.

Recent calculations as reported by I. Ivanov~\cite{iiva} and
Martin~\cite{mrt} improve on these approximations (see
also~\cite{rev1}). Firstly, the transverse momentum of the exchanged
gluons is taken into account by applying the so-called
$k_T$-factorization and using the {\em unintegrated} gluon distribution
$f(\xi,k_T^2)$, where $k_T$ is the transverse momentum of the
exchanged gluons.  Secondly, even in the case of forward scattering
the need to transform a spacelike photon into a timelike vector meson
forces the kinematics to be non-forward, and the usual parton (gluon)
distributions have to be replaced by {\em skewed} (also called
non-forward or off-diagonal) parton distribution functions (SPDF).
These are generalizations of the normal PDFs (without direct
probabilistic interpretation) and follow their own, new evolution
equations.  A method to construct corresponding exclusive
evolution kernels at NLO was reported by Freund~\cite{bfm}. In this 
work explicit diagrammatic two-loop calculations are avoided by using
conformal ${\cal N} = 1$ SUSY Yang-Mills constraints together with
known two-loop DGLAP kernels.  

Within the perturbative two-gluon-picture, elastic (electro-)
production of light and heavy vector mesons can be calculated in fair
agreement with experimental data as long as $Q^2$ and/or $M_V^2$
provide a hard scale of several GeV$^2$.  As the cross section depends
on the gluon distribution squared, the process $\gamma^* p \to V p$ may
serve as a particularly sensitive probe of the gluon at small $\xi$.
To achieve this ambitious goal the theoretical uncertainties should be
decreased further.  In addition use should be made of the different
available observables, i.e., $Q^2$ and energy dependence of the total
cross sections for different mesons, $\sigma_L/\sigma_T$, the ratio of
longitudinal to transverse photon induced production, and maybe even
the full spin density matrix of $\rho$ or $J/\psi$ production
measurements.  One particular source of uncertainty in the theoretical
description is the meson wave function.  As shown by
I. Ivanov~\cite{iiva} different wave function models lead to quite
different results, especially for $\sigma_T$, and the wave function
is expected to play a significant role in the production of excited
vector mesons compared to ground states.

On the other hand, as demonstrated by Martin~\cite{mrt}, the basic
features of elastic vector meson production are mainly controlled by the
photon wave function and the gluon distribution and can therefore be
well predicted in the framework of Parton-Hadron-Duality (PHD).  This
approach uses open $q\bar q$ pair production, integrated over an
appropriate mass interval and projected on the quantum numbers of the
meson under consideration, thus avoiding the meson wave function
completely.  Even $\sigma_L/\sigma_T$, which is poorly described in
most other models, is in agreement with HERA data.

Another field where the use of perturbative QCD in the framework of
the two gluon picture can be justified is diffractive meson
production at large momentum transfer $t$.  New interesting results
were reported by D.\ Ivanov~\cite{diva,dp}.  He obtained large
contributions to high-$t$ light vector meson photoproduction from a --
normally highly suppressed -- chiral odd $q\bar q$ component, where
the photon couples to the quarks via the magnetic susceptibility of
the vacuum with a surprisingly large coefficient.

A particularly interesting possibility for observing the odderon,
i.e., the $C=P=-1$ partner of the pomeron, in the context of diffractive meson 
production has been suggested by Dosch~\cite{do,ber}. Addressing the 
fundamental question why the odderon is not seen in the difference between 
$pp$ and $p\bar{p}$ cross sections, the authors suggest that the reason 
lies in the quark-diquark structure of the proton. If this is the case, 
then the diffractive production of pseudoscalar mesons at HERA with proton 
breakup in the final state should provide an ideal testing ground for the 
odderon. Employing a stochastic vacuum based approach in the description 
of the soft $t$-channel exchange~\cite{msv}, a prediction of $\sigma_{
\gamma p\to\pi^0X}\approx 300$ nb with estimated model uncertainties of 
$\pm 50\%$ was given.

\section{Conclusions and Outlook}
The high-energy asymptotics of hadronic cross sections belong to the few 
phenomenologically relevant and not yet calculable implications of known 
quantum field theories. Thus, high-energy scattering and diffraction remain 
among the least well understood and therefore most interesting fields in 
QCD. Progress has been reported in may directions, with work ranging from 
simple phenomenology to elaborate multi-loop calculations. Given the fast 
and continuous improvement of data and the rich interplay between theory 
and experiment, we experience an exciting time for this fundamental area of 
research.

\section*{Acknowledgements}
We would like to thank the organizers for this very interesting and 
stimulating Workshop and the participants of our Working Group for their 
excellent presentations and for their enthusiasm during the discussions.

\end{document}